\documentclass[12pt]{article}
\usepackage{graphicx}
\begin{document}

\pagestyle{empty}

\noindent
{\bf Does the Period of a Pulsating Star Depend on its Amplitude?}

\bigskip

\noindent
{\bf John R. Percy and Jeong Yeon (JY) Yook\\Department of Astronomy and Astrophysics\\University of Toronto\\Toronto ON\\Canada M5S 3H4}

\bigskip

{\bf Abstract}  Several classes of pulsating stars are now known to undergo
slow changes in amplitude; these include pulsating red giants and
supergiants, and yellow supergiants.  We have used visual observations
from the AAVSO International Database, and wavelet analysis of 39 red
giants, 7 red supergiants, and 3 yellow supergiants, to test the hypothesis
that an increase in amplitude would result in an increase in period, because
of non-linear effects in the pulsation.  For most of the stars, the
results are complex and/or indeterminate, due to the limitations of the data,
the small amplitude or amplitude variation, or other processes such as
random cycle-to-cycle period fluctuations.  For the dozen stars which have substantial 
amplitude variation, and reasonably simple behavior, there is a 75-80\% tendency to show a
positive correlation between amplitude and period.

\medskip

\noindent
{\bf 1. Introduction}

\smallskip

Galileo Galilei is noted for (among other things) observing that the period of the swing of a
pendulum does not depend on the amplitude of the swing.  For most
vibrating objects, however, there are non-linear effects which cause the period
to increase slightly if the amplitude becomes sufficiently large.

We have recently noted that there are systematic, long-term variations
in amplitude in pulsating red giants (Percy and Abachi 2013), pulsating
red supergiants (Percy and Khatu 2014), and pulsating yellow supergiants
(Percy and Kim 2014).  The purpose of this project was to investigate
whether there might be systematic changes in period which accompany the
changes in amplitude.  This possibility has already been suggested 
as occuring
in R Aql, BH Cru, and S Ori (Bedding {\it et al.} 2000, Zijlstra {\it et al.} 2004).

Our study is complicated by several factors.  Stars undergo small,
slow evolutionary changes in period.  They also undergo random cycle-to-cycle
fluctuations in period (Eddington and Plakidis 1929, Percy and Colivas 1999).
We have shown that, for some reason, the amplitudes themselves are variable.
The stars are complicated: red giants and supergiants have large,
convective hot and cool regions on their surfaces.  Furthermore, the stars
rotate with periods which are comparable with the time scales for amplitude
change.  For these reasons, it may be difficult to isolate any non-linear effect
of changing amplitude on period.

\medskip

\noindent
{\bf 2. Data and Analysis}

\smallskip

We used visual observations, from the AAVSO International Database, of the
stars listed in Tables 1-3.  See "Notes on Individual Stars" for remarks on
some of these.  Percy and Abachi (2013) discussed some of the limitations
of visual data which must be kept in mind when analyzing the observations,
and interpreting the results, but only visual observations are sufficiently
dense, sustained, and systematic for use in this project.  
The data, extending from JD(1) to JD(1) +
$\Delta$JD [JD = Julian Date, in days] as given in
the tables, were analyzed
using the VSTAR package (Benn 2013; www.aavso.org/vstar-overview),
especially the wavelet (WWZ) analysis routine.  The periods of the
stars had previously been determined with the DCDFT routine.  For the
wavelet analysis, as in our previous papers, the default values were
used for the decay time c (0.001) and time division $\Delta$t (50 days).
The results are sensitive to the former, but not to the latter.

We generated light curves and graphs of period and amplitude versus JD,
but our main tool for analysis was graphs of amplitude versus period,
as shown in the figures.  For each of these, the method of least squares
was used to determine the straight line of best fit, the slope k of
this fit, the standard error $\sigma$ of the fit, and the coefficient
of correlation R.  Tables 1-3 give the star, period in days, initial JD
and range of JD, amplitude and amplitude range in magnitudes, k, $\sigma$,
k/$\sigma$, R, and any notes.  See also the ``Notes on Individual Stars".
In the last column, an asterisk (*) indicates that k/$\sigma$ is greater
than 3, and a double asterisk (**) indicates that R $\ge$ 0.5 i.e. the
results are statistically significant.  The notation e-x means 10 to the power x.
The Notes column also includes a qualitative description
of the shape and trajectory of the semi-amplitude versus period
plots: 1 indicates positive slope,
2 indicates negative slope, 3 indicates non-linear, 4 indicates
vertical lines, 5 indicates counterclockwise, 6 indicates
clockwise, 7 indicates irregular trajectory, and 8 indicates a sinusoidal
trajectory.  The spectral types in the figure captions are from SIMBAD.
 
\medskip

\noindent
{\bf 3. Results}

\medskip

\noindent
3.1.  Red giants

\smallskip

Table 1 shows the results of the single-mode variables
from Kiss {\it et al.} (2006), Percy and Abachi (2013), Bedding {\it et al.} (2000),
and Zijlstra {\it et al.} (2004). 
In total, 39 pulsating red giants
are listed on the table and 26 of them have a positive k and 13
of them have a negative k. 

Figure 1 
shows the semi-amplitude versus period relationship for BH Cru.
It displays a strong,
positive correlation between semi-amplitude and period;
and there is almost no non-linearity.  The relation is also
clear from the plots of period-JD and amplitude-JD (Bedding {\it et al.} (2000).
Note that we are using about 14 more years of data than Bedding {\it et al.} (2000).

Figure 2 shows the semi-amplitude versus period plot
for R Aql. There is a strong, positive
correlation between them as listed in Table 1. 
This agrees with the discussion of Bedding {\it et al.} (2000)
that R Aql shows some relationship between period and
amplitude. However, there is local non-linearity in
addition to the linear correlation, which suggests that there
is also some other process which affects the period.  The individual
period and amplitude plots show that, whereas the period is decreasing
monotonically from 310 to 270 days, the amplitude is decreasing but also
undergoing fluctuations, perhaps due to stochastic excitation and decay.

Figure 3 is a
semi-amplitude versus period plot for S Ori, which is again from
Bedding {\it et al.} (2000). This one also has a positive correlation;
however, there is a global non-linearity and the linear fit
does not represent the relationship between amplitude and period very well.
In this case, the individual period and amplitude plots show that the
period is undergoing fluctuations between 405 and 440 days.

Figure 4 is for GY Aql, a pulsating red giant from
Percy and Abachi (2013). The semi-amplitude and the period of GY
Aql have a sinusoidal relationship and the linear fit is not
a good representation of the data. This plot is non-linear;
however, this sinusoidal pattern shows more regularity than
other non-linear plots.  Note, however, that the change in
amplitude is small, both absolutely and as a fraction of the
average amplitude. 

Figure 5 is for S Aur from
Percy and Abachi (2013). There is a positive correlation between the
semi-amplitude and period, but this obviously not the dominant process
affecting the period.

Figure 6 shows
the semi-amplitude versus period for S Cam. There is a positive
correlation with some non-linearity.  The change in amplitude is relatively
small.

Figure 7 is for DM
Cep and it has a negative slope with some non-linearity. The data
has a gap in the mid-region of the data, and the amplitude is very small. 

Figure 8 is
for SY Per and there is a positive correlation. There is a local
non-linearity in the plot; however, the line of the best fit is
a good representation of the data globally. 

Figure 9 
is for UZ Per. UZ Per has a long period and a small change in
amplitude, so the negative slope is not really meaningful.

Figure
10 is for W Tau. There is a positive correlation between
semi-amplitude and period, with some non-linearity in the plot
in addition to the global linear trend.

\begin{figure}
\begin{center}
\includegraphics[height=6cm]{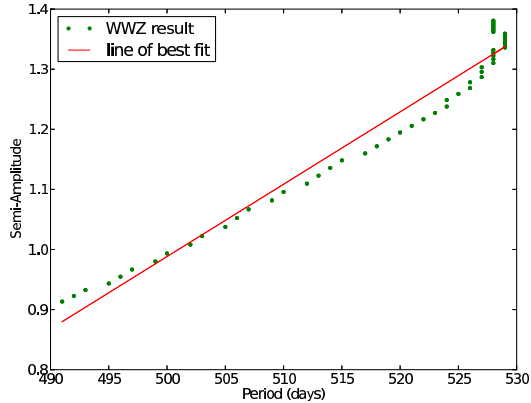}
\end{center}
\caption{Amplitude versus period for BH Cru (SC4.5-7/8e).  The correlation is
excellent, as was apparent from the graphs of period and amplitude
versus JD (Bedding {\it et al.} 2000)}
\end{figure}

\begin{figure}
\begin{center}
\includegraphics[height=6cm]{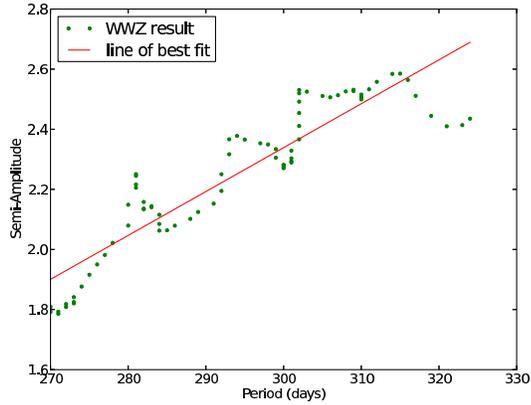}
\end{center}
\caption{Amplitude versus period for R Aql (M6.5-9e).  The positive correlation
was suggested by Bedding {\it et al.} (2000) on the basis of the graphs of period
and amplitude versus JD.  The deviations from the straight-line fit
suggest that there is one or more additional factors which affect the period
and/or amplitude.}
\end{figure}

\begin{figure}
\begin{center}
\includegraphics[height=6cm]{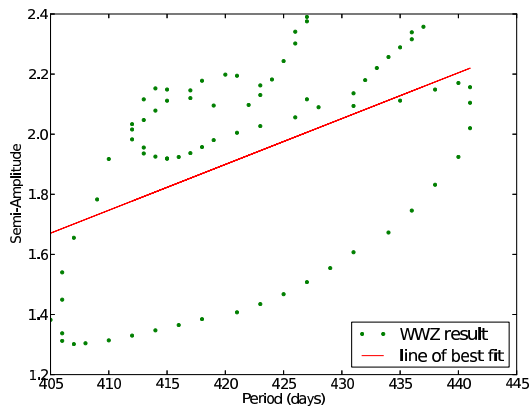}
\end{center}
\caption{Amplitude versus period for S Ori (M6.5-7.5e).  A positive correlation
was suggested by Bedding {\it et al.} (2000) but it is clear that, although this graph
shows such a correlation, the correlation is weak, presumably because of other processes which affect the period and/or amplitude.}
\end{figure}

\begin{figure}
\begin{center}
\includegraphics[height=6cm]{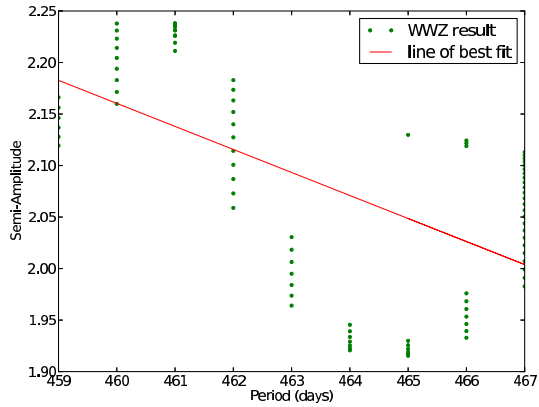}
\end{center}
\caption{Amplitude versus period for GY Aql (M6e).  The correlation is negative 
but the change in amplitude and period is very small, relative to their mean values.}
\end{figure}

\begin{figure}
\begin{center}
\includegraphics[height=6cm]{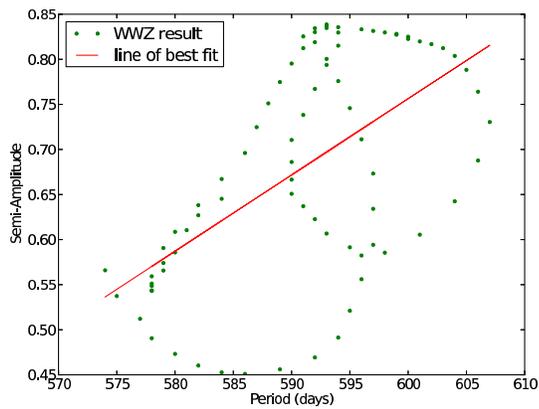}
\end{center}
\caption{Amplitude versus period for S Aur (N0).  The correlation is positive
but weak and non-linear, indicating that other factors are important in determining
the changes in period and/or amplitude.}
\end{figure}

\begin{figure}
\begin{center}
\includegraphics[height=6cm]{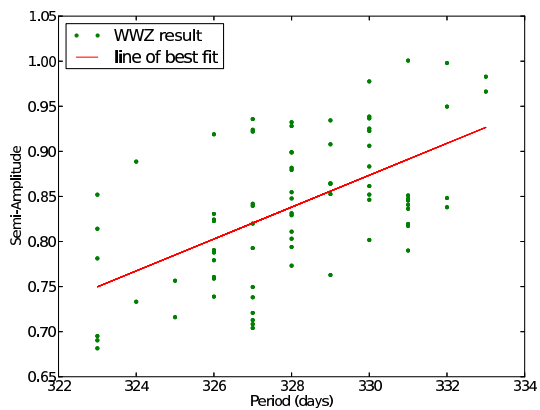}
\end{center}
\caption{Amplitude versus period for S Cam (R8e).  The correlation is positive
but scattered and weak.}
\end{figure}

\begin{figure}
\begin{center}
\includegraphics[height=6cm]{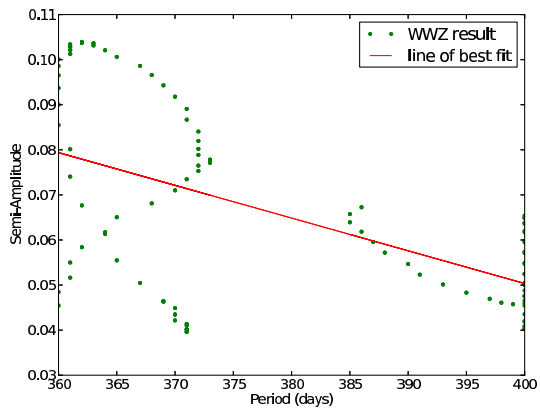}
\end{center}
\caption{Amplitude versus period for DM Cep (M3D).  The small amplitude makes
any correlation meaningless.}
\end{figure}

\begin{figure}
\begin{center}
\includegraphics[height=6cm]{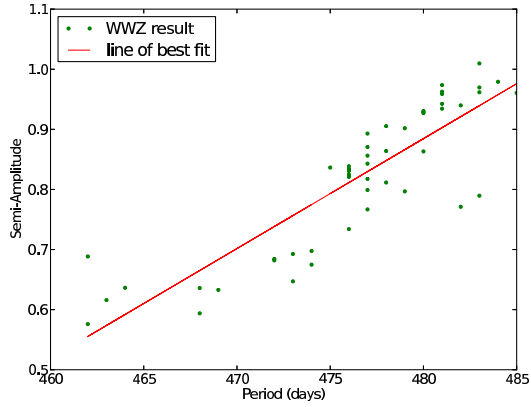}
\end{center}
\caption{Amplitude versus period for SY Per (C6,4e).  There is a significant
positive correlation, with some deviations from this.}
\end{figure}

\begin{figure}
\begin{center}
\includegraphics[height=6cm]{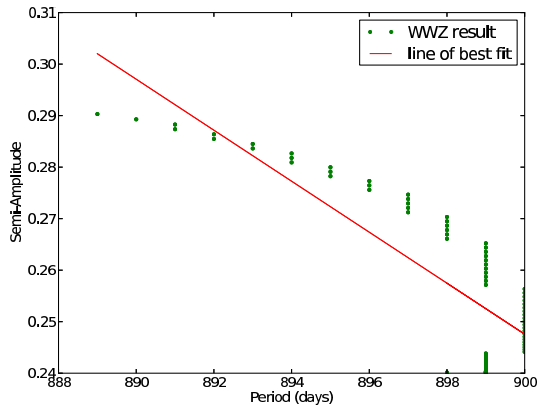}
\end{center}
\caption{Amplitude versus period for UZ Per (M5II-III).  The
correlation is negative, but the amplitude and its change is very small.}
\end{figure}

\begin{figure}
\begin{center}
\includegraphics[height=6cm]{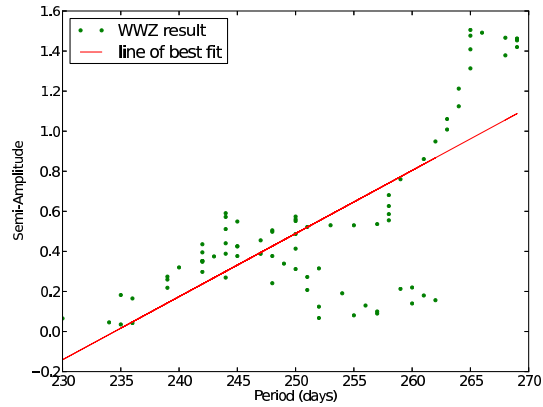}
\end{center}
\caption{Amplitude versus period for W Tau (M6D).  There is a positive
correlation, with some deviations.  The change in period is exceptionally
large -- 15 per cent.}
\end{figure}

\begin{figure}
\begin{center}
\includegraphics[height=6cm]{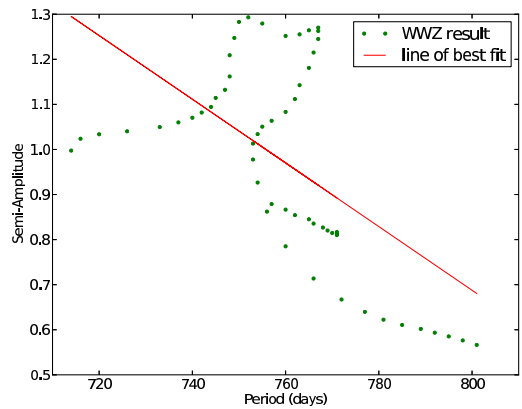}
\end{center}
\caption{Amplitude versus period for VX Sgr (M5/6III or M4Iae).  There
is a negative correlation, but the relation is certainly not linear.  This
is not surprising in a star as complex as a red supergiant.  The
change in period is large -- 10 per cent.}
\end{figure}

\begin{figure}
\begin{center}
\includegraphics[height=6cm]{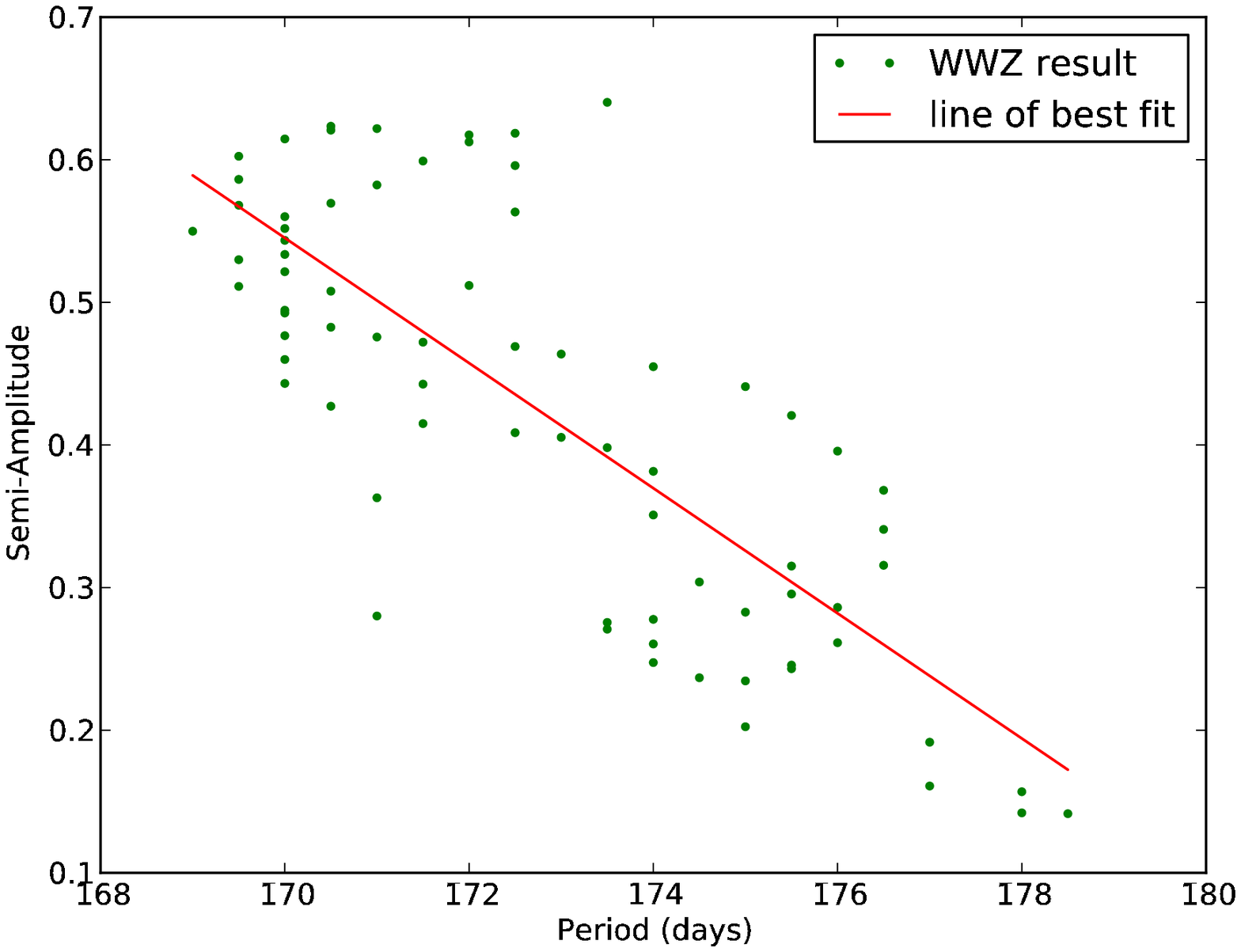}
\end{center}
\caption{Amplitude versus period for DE Her (K0D), a yellow semi-regular
variable star.  There is a strong negative correlation, with
significant deviations from this.}
\end{figure}



\medskip

\noindent
3.2. Red Supergiants

\smallskip

Table 2 presents the results of WWZ analysis for
red supergiants. The notations used are the same as in
Table 1. Seven red supergiants were studied. One had a
negative correlation and six had a positive correlation between
the semi-amplitude and the period.

Figure 11 is the semi-amplitude versus period plot for
VX Sgr. VX Sgr has a long period which is a characteristic of
supergiants.  There is some negative correlation. However, the
plot is non-linear and the line of best fit does not represent
the plot well.  This is not surprising, in view of the complexity
of this class of stars.

\begin{figure}
\begin{center}
\includegraphics[height=5cm]{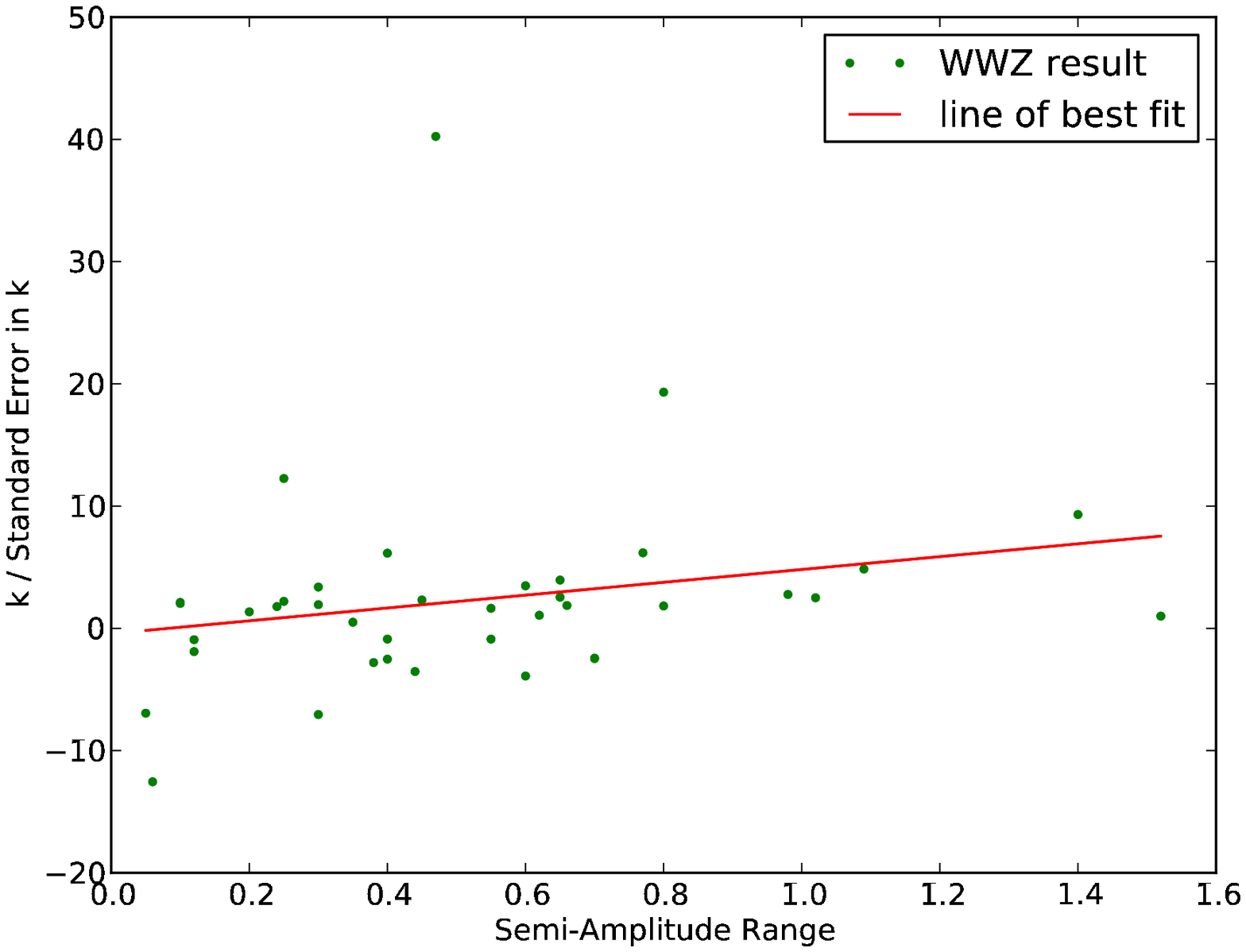}
\end{center}
\caption{The relationship between k/$\sigma$ and the range in amplitude.
There is a slight positive correlation.}
\end{figure}

\medskip

\noindent
3.3. Yellow Supergiants

\smallskip

Table 3 displays the result of WWZ analysis for
yellow supergiants. The notations used are the same as in
Table 1. Three yellow supergiants were studied and they
all have a non-linear relationship between the semi-amplitude
and the period, but only one is significant.

Figure 12 is the semi-amplitude versus period plot for DE
Her. There is a weak negative correlation. The plot is non-linear
and the line of best fit does not describe the plot in a meaningful way.

\medskip

\noindent
3.4.  Summary Statistics

\smallskip

Figure 13
is a plot for $k/\sigma$ versus amplitude
range. The slope of the linear fit was 0.19. The standard error in
the slope was 3.29. The coefficient of correlation was 0.53. In
both plots, the correlation is positive. It is not a strong
correlation, but not weak either.

\begin{center}
\begin{table}
\footnotesize
\caption{Variability and WWZ results of pulsating red giants.}
\begin{tabular}{l*{7}{c}{c}{c}{c}r}
\hline
Star & P(d) & P Range & JD(1) & $\Delta$JD & A & A Range & k & $\sigma$ & $k/\sigma$ & $R$ & Notes \\
\hline\hline
RV And & 165 & 38 & 2428000 & 28300 & 0.30 & 0.20-0.60 & -2.54e-3 & 2.9e-3 & -0.88 & 1.2e-1 & 3, 7\\
RY And & 392 & 11 & 2427500 & 30000 & 1.69 & 0.68-2.20 & 1.89e-2 & 1.9e-2 & 1.00 & 1.3e-1 & 4, 7\\
R Aql & 294 & 54 &2415000 & 40000 & 2.25 & 1.79-2.59 & 1.46e-2 & 7.6e-4 & 19.32 & 9.1e-1 & **, 1, 7\\
S Aql & 143 & 8 & 2420000 & 36300 & 0.98 & 0.65-1.20 & -1.45e-2 & 1.7e-2 & -0.88 & 1.0e-1 & 4, 7 \\
GY Aql & 464 & 8 & 2447000 & 9300 & 2.35 & 1.90-2.20 & -2.23e-2 & 3.2e-3 & -7.05 & 6.0e-1 & **, 8  \\
RS Aqr & 280 & 8 & 2430000 & 27500 & 2.73 & 2.58-2.97 & -2.00e-2 & 7.2e-3 & -2.80 & 3.6e-1 & 4, 7 \\
T Ari & 320 & 13 & 2428000 & 28300 & 0.91 & 0.70-1.35 & 1.59e-2 & 6.2e-3 & 2.55 & 3.2e-1 & 3, 5 \\
S Aur & 596 & 33 & 2416000 & 40300 & 0.61 & 0.45-0.85 & 8.46e-3 & 1.4e-3 & 6.15 & 5.7e-1 & **, 3, 7 \\
U Boo & 204 & 10 & 2420000 & 49300 & 0.62 & 0.35-0.80 & 1.62e-2 & 7.0e-3 & 2.32 & 2.6e-1 & 3, 7 \\
V Boo & 887 & 50 & 2415000 & 40000 & 0.31 & 0.06-0.50 & -3.43e-3 & 9.7e-4 & -3.53 & 3.7e-1 & *, 3, 5 \\
RV Boo & 144 & 29 & 2434000& 22300 & 0.09 & 0.05-0.15 & 1.16e-3 & 5.7e-4 & 2.04 & 1.9e-1 & 3, 7  \\
S Cam & 327 & 10 & 2417000 & 39300 & 0.34 & 0.23-1.00 & 1.77e-2 & 2.9e-3 & 6.18 & 5.7e-1 &  **, 3, 7 \\
RY Cam & 134 & 6 & 2435000 & 21300 & 0.16 & 0.10-0.40 & 1.99e-2 & 5.9e-3 & 3.38 & 3.1e-1 & *, 4, 7 \\
T Cnc & 488 & 19 & 2417000 & 39300 & 0.34 & 0.23-0.47 & 2.89e-3 & 1.6e-3 & 1.78 & 2.0e-1 & 3, 7 \\
RT Cap & 400 & 75 & 2417000 & 39300 & 0.31 & 0.25-0.45 & 4.52e-4 & 3.4e-4 & 1.35 & 4.5e-4 & 3, 7 \\
T Cen & 91 & 19 & 2413000 & 43300 & 0.62 & 0.50-1.20 & -4.51e-2 & 1.8e-2 & -2.47 & 2.6e-1 & 4, 7 \\
DM Cep & 367 & 40 & 2435000 & 21300 & 0.12 & 0.05-0.10 & -7.26e-4 & 1.0e-4 & -6.94 & 5.6e-1 &  **, 4, 7 \\
T CMi & 321 & 23 & 2415000 & 40000 & 1.86 & 1.24-2.27 & 1.25e-2 & 5.0e-3 & 2.50 & 2.7e-1 & 4, 5 \\
RS CrB & 331 & 7 & 2435000 & 21300 & 0.19 & 0.13-0.38 & 7.79e-3 & 3.5e-3 & 2.21 & 2.1e-1 & 5, 7 \\
BH Cru & 518 & 38 & 2440000 & 10000 & 1.21 & 0.91-1.38 & 1.20e-2 & 3.0e-4 & 40.24 & 9.8e-1& **, 1 \\
T CVn & 291 & 15 & 2415000 & 40000 & 0.83 & 0.57-1.27 & -1.59e-2 & 6.5e-3 & -2.44 & 6.5e-3 & 3, 7 \\
RU Cyg & 234 & 14 & 2415000 & 40000 & 0.38 & 0.11-0.77 & 1.33e-2 & 7.1e-3 & 1.87 & 2.1e-1 & 3, 7 \\
V460 Cyg & 160 & 15 & 2435000 & 21300 & 0.08 & 0.04-0.14 & 2.51e-3 & 1.2e-3 & 2.12 & 2.0e-1 & 3, 7 \\
V930 Cyg & 247 & 43 & 2442000 & 14300 & 0.72 & 0.30-0.70 & -5.83e-3 & 2.3e-3 & -2.52 & 2.9e-1 & 3, 7 \\
EU Del & 62 & 25 & 2435000 & 21300 & 0.08 & 0.05-0.17 & -1.20e-3 & 1.3e-3 & -0.93 & 9.0e-2 & 3, 7 \\
SW Gem & 700 & 117 & 2427500 & 28800 & 0.10 & 0.05-0.35 & 1.55e-3 & 8.0e-4 & 1.94 & 2.5e-1 & 3, 7 \\
RR Her & 250 & 12 & 2435000 & 21300 & 0.54 & 0.10-0.70 & 1.33e-2 & 3.8e-3 & 3.48 & 3.2e-1& *, 3, 5 \\
RT Hya & 255 & 29 & 2415000 & 41300 & 0.06 & 0.04-0.16 & 9.30e-3 & 5.1e-3 & 1.83 & 2.0e-1 & 3, 7 \\
U Hya & 791 & 98 & 2420000 & 36300 & 0.06 & 0.04-0.16 & -2.22e-4 & 1.2e-4 & -1.90 & 2.2e-1 & 3, 7 \\
U LMi & 272 & 31 & 2427500 & 30000 & 0.50 & 0.23-0.85 & 3.98e-3 & 3.7e-3 & 1.07 & 1.4e-1 & 3, 7 \\
X Mon & 148 & 9 & 2415000 & 41300 & 0.59 & 0.25-0.85 & -3.04e-2 & 7.8e-3 & -3.90 & 4.0e-1 & *, 4, 7 \\
S Ori & 422 & 36 & 2415000 & 40000 & 1.93 & 0.30-2.39 & 1.52e-2 & 3.1e-3 & 4.85 & 4.9e-1 & *, 3, 5 \\
S Pav & 387 & 11 & 2415000 & 40000 & 0.70 & 0.30-1.29 & 2.38e-2 & 8.6e-3 & 2.77 & 2.9e-1 & 3, 7 \\
Y Per & 251 & 11 & 2415000 & 40000 & 0.72 & 0.34-0.99 & 2.49e-2 & 6.3e-3 & 3.95 & 4.1e-1 & *, 3, 7 \\
SY Per & 477 & 23 & 2446000 & 10300 & 0.89 & 0.67-0.92 & 1.83e-2 & 1.5e-3 & 12.26 & 8.7e-1 & **, 1 \\
UZ Per & 850 & 11 & 2448000 & 8300 & 0.25 & 0.23-0.29 & -4.95e-3 & 3.9e-4 & -12.55 & 8.1e-1 & **, 2 \\
W Tau & 243 & 39 & 2415000 & 41300 & 0.27 & 0.10-1.50 & 3.15e-2 & 3.4e-3 & 9.31 & 7.2e-1 & **, 1 \\
V UMa & 198 & 42 & 2420000 & 36300 & 0.19 & 0.15-0.50 & 5.71e-4 & 1.1e-3 & 0.50 & 6.0e-2 & 3, 7 \\
SS Vir & 361 & 17 & 2420000 & 36300 & 0.82 & 0.60-1.15 & 4.56e-3 & 2.8e-3 & 1.64 & 1.9e-1 & 3, 5 \\
\hline
\end{tabular}
\end{table}
\end{center}

\begin{center}
\begin{table}
\small
\caption{Variability and WWZ results of pulsating red supergiants.}
\begin{tabular}{l*{7}{c}{c}{c}{c}r}
\hline
Star & P(d) & P Range & JD(1) & $\Delta$JD & A & A Range & k & $\sigma$ & $k/\sigma$ & $R$ & Notes \\
\hline\hline

BO Car & 337 & 20 & 2443000 & 14000 & 0.13 & 0.07-0.21 & -5.6e-3 & 1.3e-3 & -4.45 & 4.5e-1 & *, 2, 7 \\
PZ Cas & 846 & 24 & 2440000 & 15000 & 0.24 & 0.13-0.50 & 6.5e-3 & 1.4e-3 & 4.66 & 4.6e-1 & *, 3, 5 \\
BC Cyg & 703 & 25 & 2440000 & 15000 & 0.30 & 0.14-0.51 & 7.6e-3 & 2.1e-3 & 3.71 & 3.8e-1 & *, 3, 5 \\
W Ind & 194 & 45 & 2443000 & 14000 & 0.40 & 0.95-1.09 & -9.0e-4 & 1.4e-3 & -0.63 & 7.0e-2 & 3, 7 \\
S Per & 809 & 44 & 2420000 & 35000 & 0.57 & 0.33-0.85 & -9.1e-5 & 1.4e-3 & -0.06 & 7.5e-3 & 3, 7 \\
W Per & 489 & 87 & 2415000 & 40000 & 0.35 & 0.19-0.48 & -1.0e-4 & 3.3e-4 & -0.31 & 3.4e-2 & 3, 7 \\
VX Sgr & 760 & 87 & 2427500 & 30000 & 0.73 & 0.57-1.30 & -7.1e-3 & 1.3e-3 & -5.36 & 5.8e-1& **, 3, 7 \\

\hline
\end{tabular}
\end{table}
\end{center}

\begin{center}
\begin{table}
\small
\caption{Variability and WWZ results of pulsating yellow supergiants.}
\begin{tabular}{l*{7}{c}{c}{c}{c}r}
\hline
Star & P(d) & P Range & JD(1) & $\Delta$JD & A & A Range & k & $\sigma$ & $k/\sigma$ & $R$ & Notes \\
\hline\hline

AV Cyg & 88 & 5 & 2430000 & 27500 & 0.37 & 0.12-0.56 & -1.0e-2 & 1.4e-2 & -0.73 & 9.9e-2 & 3, 7 \\
DE Her & 173 & 10 & 2442500 & 12500 & 0.42 & 0.14-0.64 & -4.4e-2 & 4.1e-3 & -10.77 & 7.9e-1 & **, 3, 7 \\
RS Lac & 238 & 2 & 2427500 & 30000 & 0.72 & 0.35-1.03 & 7.4e-2 & 5.3e-2 & 1.40 & 1.8e-1 & 4, 7 \\

\hline
\end{tabular}
\end{table}
\end{center}

\medskip

\noindent
{\bf 3.4. Notes on Individual Stars}.

\smallskip

{\it RY And}: There are some sparse regions of data in between dense regions.

\smallskip

{\it GY Aql}: The data are sparse. 

\smallskip

{\it R Aql}: The data are sparse before 2420000. There is an outlier in period in the beginning.

\smallskip

{\it S Aql}: There is an abrupt change in period at the end of the data.

\smallskip

{\it RV Boo}: The period is not smooth.

\smallskip

{\it RY Cam}: The period is not smooth.

\smallskip

{\it RT Cap}:  The data are sparse near JD = 2430000. There is an abrupt change in period in the middle.

\smallskip

{\it BO Car}: The data are sparse before JD = 2443000.

\smallskip

{\it T Cen}: The period is not smooth. The data are sparse near JD = 2430000. There is an abrupt change in period in the middle.

\smallskip

{\it DM Cep}: The data are sparse between JD = 2440000 and JD = 2442500. There is an abrupt change in period in the middle.

\smallskip

{\it V460 Cyg}: The period is not smooth.

\smallskip

{\it V930 Cyg}: The data are sparse before JD = 2445000. There is an outlier in period in the beginning.

\smallskip

{\it EU Del}: The period and the semi-amplitude are not smooth. There are two outliers in the light curve. 

\smallskip

{\it SW Gem}: There is an abrupt change in the middle.

\smallskip

{\it RR Her}: The period and the semi-amplitude are not smooth.

\smallskip

{\it RT Hya}: There is an abrupt change in period. The semi-amplitude is not smooth.

\smallskip

{\it U Hya}: The data are sparse near JD = 2430000 There is an abrupt change in period in the middle.

\smallskip

{\it W Ind}: The data are sparse near JD = 2452500. There is an abrupt change in period and amplitude at the end.

\smallskip

{\it U LMi}:  There is an abrupt change of period at the end.

\smallskip

{\it X Mon}: The period and the semi-amplitude are not smooth.

\smallskip

{\it S Ori}: The data are sparse before 2420000.

\smallskip

{\it S Pav}: The data are sparse from JD = 2420000 to JD = 2427000.

\smallskip

{\it S Per}: The data are sparse before JD =  2420000.

\smallskip

{\it SY Per}: The data are sparse before JD = 2448000.

\smallskip

{\it VX Sgr}: There is an abrupt change in period in the beginning.

\smallskip

{\it W Per}: There is an abrupt change in period in the beginning.

\smallskip

{\it W Tau}: There is an abrupt change in period in the middle.

\medskip

\noindent
{\bf 4. Discussion}

\smallskip

There are a variety of mechanisms which could cause period (or amplitude) changes in
pulsating red giants and supergiants, and other cool, luminous stars: evolution,
random cycle-to-cycle fluctuations, helium shell flashes, or simply
the complexity of a star with large convective cells which is rotating
and losing mass.  Nevertheless: if we restrict our attention to stars
whose amplitude and amplitude changes are sufficiently large, and whose
amplitude versus period relation has a statistically significant linear
slope, then 9 of 11 pulsating red giants show a period which increases
with increasing amplitude.  Choosing slightly differently: among stars
with amplitudes greater than 1.0 mag, and significant {\it changes} in
amplitude, 10 of 12 have a positive correlation between amplitude and
period.  This is not to say, of course, that the period change is
{\it caused} by the amplitude change.

We must also remember that the visual light curve is not a bolometric
light curve and that, for red stars, the visual band is especially
sensitive to temperature, which may not have a direct effect on the
pulsation period.

\medskip

\noindent
{\bf 5. Conclusions}

\smallskip

In stars with a variable pulsation amplitude, does an increase in pulsation amplitude result in an increase in period?
The majority of the almost-50 pulsating stars in our sample do {\it not}
show a linear relation between the instantaneous period and amplitude.
Clearly, there are other processes which affect the period and amplitude.
But, of the dozen stars which show sufficiently large amplitude and
amplitude change, 75-80\% show a positive correlation between the
instantaneous amplitude and period.

\medskip

\noindent
{\bf Acknowledgements}

\smallskip

We thank the hundreds of AAVSO observers who made the observations which were
used in this project, and we thank the AAVSO staff for processing and
archiving the measurements.  We also thank the team which developed the
{\it VSTAR} package, and made it user-friendly and publicly available.  We
are especially grateful to Professor Tim Bedding for suggesting this
project in the first place.
This project
made use of the SIMBAD database, which is operated by CDS,
Strasbourg, France.

\medskip

\noindent
{\bf References}

\smallskip

\noindent
Bedding, T.R., Conn, B.C., and Zijlstra, A.A., 2000, in {\it The Impact of
Large-Scale Surveys on Pulsating Star Research}, eds. L. Szabados and D.W. Kurtz,
ASP Conf. Ser. 203, Astronomical Society of the Pacific, San Francisco, 96.

\smallskip

\noindent
Benn, D. 2013, VSTAR data analysis software (http://www.aavso.org/node/803).

\smallskip

\noindent
Eddington, A.S. and Plakidis, S., 1929, {\it Mon. Not. Roy. Astron. Soc.}, {\bf 90}, 65.
 
\smallskip

\noindent
Kiss, L.L., {\it et al.}, 2006, {\it Mon. Not. Roy. Astron. Soc.}, {\bf 372}, 1721.

\smallskip

\noindent
Percy, J.R. and Colivas, T., 1999, {\it Publ. Astron. Soc. Pacific}, {\bf 111}, 94.

\smallskip

\noindent
Percy, J.R. and Abachi, R., 2013, {\it JAAVSO}, {\bf 41}, 193. 

\smallskip

\noindent
Percy, J.R., and Khatu, V., 2014, {\it JAAVSO}, in press.

\smallskip

\noindent
Percy, J.R., and Kim, R., 2014, http://arxiv.org/abs/1405.6993

\smallskip

\noindent
Zijlstra, A.A., {\it et al.}, 2004, {\it Mon. Not. Roy. Astron. Soc.}, {\bf 352}, 325.

\end{document}